# Retinal processing of natural scenes : challenges ahead

Samuele Virgili[1,2], Olivier Marre[1,2]

1 Institut de la Vision, Sorbonne Université, INSERM, CNRS, Paris, France
2 Corresponding authors : samuele.virgili@yahoo.com, olivier.marre@gmail.com

## Abstract

While a great deal is known about the way the retina processes simple stimuli, our understanding of how the retina processes natural stimuli is still limited. Here we highlight some of the challenges that remain to be addressed to understand retinal processing of natural stimuli and describe emerging research avenues to overcome them.

A key issue is model complexity. When complexifying the probing stimuli towards natural stimuli, the number of parameters required in models of retinal computations increases, raising issues of overfitting, generalization, and interpretability. This increase in complexity is also a challenge for normative approaches as it makes it difficult to derive non-linear retinal computations from simple principles.

We describe two types of approaches that may help circumvent this issue in the future. First, we propose that a new form of reductionism is emerging: instead of breaking down natural stimuli into sums of simpler stimuli, it becomes possible to "divide and conquer" natural scenes into different visual inputs corresponding to different visual tasks, in order to study retinal computations separately for each of these visual tasks. Second, several studies suggest that it will soon be possible to mitigate the issue of complexity, by "embodying" the models with more biological constraints, in particularly those derived from connectomic studies.
Together, these approaches may offer a powerful strategy to move beyond current limitations and advance our understanding of how the retina processes natural visual environments, and suggest approaches that could be used beyond, for other sensory areas.

1. Introduction

Much of what is known about retinal processing comes from experiments using simple, artificial stimuli. These studies have uncovered many of the circuit mechanisms and computations performed by retinal ganglion cells. However, how the retina encodes complex, naturalistic inputs remains less understood.

Recent work using natural images and movies has begun to address this gap, both through encoding models that predict neural responses and normative models based on principles such as efficient coding. While these approaches have led to progress, they also face limitations—particularly in interpretability, generalization, and behavioral relevance.

In this review, we highlight key challenges in modeling retinal responses to natural scenes and propose new directions to address them. We focus on how both encoding and normative approaches can be improved by introducing additional constraints from connectomics, behavior, and task structure. In particular, we advocate for a reductionist, task-driven framework, in which retinal computations are understood in relation to specific visually guided behaviors. This perspective can help refine models, constrain their complexity, and clarify the functional role of different cell types.

We do not aim to provide a comprehensive summary of past works as this has been done in recent reviews (Turner et al. 2019; Kerschensteiner 2022; Karamanlis, Schreyer, and Gollisch 2022; Młynarski 2025, in press). Instead, we highlight emerging trends, with an emphasis on functional and modeling perspectives. While our focus is on the retina, the issues and approaches discussed here generalize to other sensory systems. We cite studies outside the retina field where relevant, without aiming to be exhaustive.

## 2. Encoding approaches: the curse of model complexity

### 2.1. Tuning curves

Classically, one approach to studying encoding in sensory systems has been to use simplified artificial stimuli, such as spots of light, moving bars, or drifting gratings. These stimuli are parameterized by one or two parameters, allowing for a systematic analysis of neural responses as a function of those parameters. By plotting a neuron's response as a function of a given stimulus parameter, one can construct a tuning curve. If the tuning curve is flat, the neuron's firing rate remains invariant to changes in the parameter, indicating that it does not encode information related to that feature. Conversely, if the tuning curve exhibits a peak at a specific value, the neuron is deemed selective for that feature. For example, direction-selective retinal ganglion cells are considered to encode the direction of motion, as they respond robustly to stimuli moving in a particular (preferred) direction while showing little to no response to motion in the opposite (null) direction (Barlow and Hill 1963). Although intuitively appealing, this approach is inherently limited in its predictive power. Typically, one must first hypothesize that a feature is encoded, possibly based on behavioral relevance (Lettvin et al. 1959) and then test whether it is actually encoded by the system through electrophysiological recordings. Doing the opposite, i.e. inferring the extracted feature directly from neural responses to a range of complex stimuli is more challenging. It requires to have an encoding model that can predict the responses to these complex stimuli, and then to analyze the model to find if a simple, low dimension description can explain most of the response (Goldin et al. 2022).

### 2.2. Receptive fields

A more agnostic and systematic approach to studying encoding without having a pre-conceived feature in mind is to characterize a neuron's functional receptive field (Chichilnisky 2001; Vlasits 2025). Rather than imposing a low-dimensional representation on the stimulus itself, this method constrains the stimulus-response function of the neuron. It is typically assumed that the neuronal response $r$ can be approximated as a linear or quasi-linear projection of the stimulus $X$ onto a filter $K$ (i.e. the neuron's receptive field), such that $r = \phi(KX)$ where $\phi$ is a threshold nonlinearity that makes the predicted response non-negative (Figure 1, left column).

Such models are often referred to as Linear-Nonlinear (LN) models. By estimating $K$ from experimental data, one can identify the single direction in stimulus space that best describes the neuron's response. Importantly, $K$ is expressed in the same space as the stimuli (e.g., pixel values for visual images), allowing for direct interpretation of the features it extracts. In the retina, for example, $K$ often exhibits a center-surround structure. Different types of ganglion cells have receptive fields that vary in size, polarity, and relative strength between center and surround components. This approach, however, has two significant limitations. First, due to sensory adaptation, the inferred filter $K$ varies depending on the statistical properties of the stimulus ensemble used for estimation (Wienbar and Schwartz 2018). Second, the assumption of quasi-linearity, while sometimes effective for describing responses to artificial stimuli (Berry and Meister 1998; Berry et al. 1999; Pillow et al. 2008), fails to generalize to natural scenes already in the retina (Heitman et al. 2016; Vystrčilová et al. 2024). In general, quasilinear models work best for stimuli that are either spatially uniform (Berry and Meister 1998), or for specific cell types. But increasing the complexity of the stimulus usually taps into nonlinear mechanisms.

This suggests that natural stimuli engage circuit mechanisms that are not captured by linear or quasi-linear models and may reveal neural computations that remain hidden under simple artificial stimulation paradigms (Rieke and Rudd 2009).

### 2.3. Nonlinear modeling

In order to predict sensory neurons responses to natural scenes, nonlinear models are needed. In general, these models take the form

$$r = \phi\left(\sum_c K_c \psi_c(x)\right)$$

where $\psi_c(x)$ are non-linear quantities extracted from the stimulus $X$ and $K$ is a linear readout across these c quantities (Figure 1, right column).

In the retina, spatial integration for example is often well modeled by a two-layer network. Several studies have shown that ganglion cells can detect details at a scale finer than their receptive field center (Enroth-Cugell and Robson 1966; Demb et al. 1999; Schwartz et al. 2012; Grimes et al. 2014; Turner and Rieke 2016; Krieger et al. 2017; Karamanlis and Gollisch 2021). In many such cases, linear models are insufficient to explain the results, at least for some cell types. To capture responses to flashed, static natural images, a common modeling approach consists of an initial stage of linear integration, followed by rectification, then a second stage of linear integration, and finally a nonlinearity at the output. Such cascade models have been successfully applied in multiple studies (Grimes et al. 2014; Turner and Rieke 2016; Karamanlis and Gollisch 2021; Zapp et al. 2022; Karamanlis et al. 2024).

However, to model ganglion cell responses to spatio-temporal natural scenes, two layers may not be enough. Network of at least three layers seem necessary (McIntosh et al. 2016; Maheswaranathan et al. 2023; Gogliettino et al. 2024). For this reason, Deep Neural Networks (DNNs) have emerged as the currently most successful class of models in capturing neural responses, both in the retina (McIntosh et al. 2016; Goldin et al. 2022; Maheswaranathan et al. 2023; Gogliettino et al. 2024) and in the visual cortex (Cadena et al. 2019; Klindt et al. 2018) - although other types of models can achieve comparable predictive performance in some cases (Goldin et al. 2023). In DNNs, the nonlinear transformations $\psi_c(x)$, often referred to as channels, are implemented via successive layers of linear convolutions interleaved with nonlinear activation functions. The number of layers can range from a single stage to dozens, and different architectures may include anywhere from a few to hundreds of channels.

These architectural properties make DNNs easy to build and train and, most importantly, due to the high number of their parameters, highly flexible in learning complex mappings between inputs and outputs. Here, flexibility refers to the ability of these models to approximate a wide variety of functions. However, while this flexibility makes DNNs a powerful machine learning tool, it also poses a challenge for neuroscience, as it comes at the expense of model interpretability.

### 2.4. The main problems of flexibility: Overfitting and lack of Interpretability

A model with high flexibility may achieve the same task through multiple distinct parameterizations. In the context of sensory encoding addressed here, this means that for a given neural response $r$ and a fixed number of channels c, there exists a large number of possible transformations $\psi_c(x)$ and weight parameters $K$ that yield equivalent outcomes. This degeneracy raises several issues:

1) It makes the models prone to overfitting—that is, fitting noise or spurious correlations in the training data rather than capturing generalizable structure in the stimulus-response relationship. To solve overfitting and enable DNNs to generalize beyond their training set, regularization is necessary. Different regularization strategies introduce specific inductive biases into the model, which influence the class of solutions to which the model will converge. However, given the complexity of modern DNNs, it is often difficult to explicitly relate a particular regularization to the inductive bias it induces. Even with regularization, generalization can remain limited (Vystrčilová et al. 2024).

2) It complicates the task of extracting mechanistic insights from model parameters. Since multiple distinct sets of nonlinear channels can produce identical response predictions, it becomes ambiguous to determine whether or not the components of the model correspond to the actual mechanisms employed by the biological system (Bowers et al. 2023). For example, Tanaka et al (Tanaka et al. 2019) found that a deep network could predict how ganglion cells respond to an omitted stimulus in a sequence of flashes (Schwartz et al. 2007). By decomposing the contribution of the different components of their model, they speculated that this omitted stimulus response might emerge thanks to the convergence of different types of bipolar cells, with different temporal properties, onto the same ganglion cell. Later work (Ebert et al. 2024) showed instead that the key component to generate an omitted stimulus response with a correct timing is an amacrine cell with depressing inhibitory synapses. This example shows how mechanistic insights from a deep network can be misleading, even when the model reproduces the phenomenon to explain.

3) It complicates the functional interpretation of the model. If a cell can be predicted by an LN model, it is relatively easy to understand what it does: the filter of the LN model represents the "preferred" stimulus, and we know that any change orthogonal to this preferred stimulus will not change the response. However, while a classical assumption is that creating a model equates to understanding ("Verum Facturm", Vico, 1744), this may not be the case with the complex models produced by machine learning. The more complex the model becomes, the more difficult it is to interpret what it does. While some visualization techniques can help finding a low-dimension description of the functional selectivity of a neuron (Goldin et al. 2022; Höfling et al. 2024), there is no guarantee that such a low dimension description exists.

In summary, while there has been tremendous progress in modeling ganglion cell responses to more and more complex stimuli, the downside is that this has led to an increasing complexity of the models used, creating issues like overfitting and lack of interpretability. A possible way to overcome this issue has been the long-standing hypothesis that this complexity can be explained by a small set of fundamental principles (Bialek 2012).

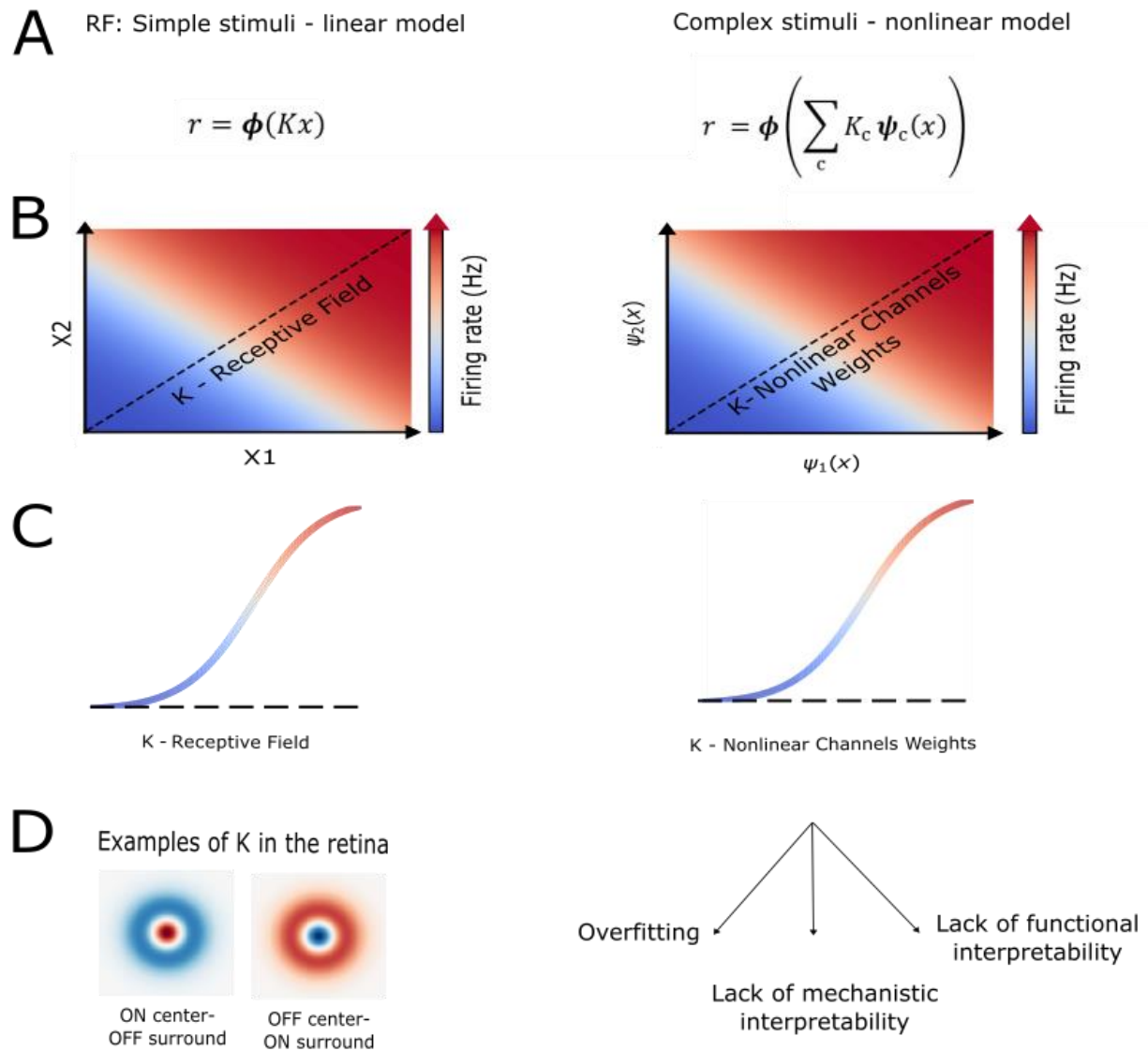

**Figure 1: Encoding approaches: the curse of model complexity: A.** Typical model structure. Left: Quasi-linear models; Right: Nonlinear models. **B.** Two dimensional stimulus-response plots. For the linear model the axis are actual stimuli, while for the nonlinear model the axis are nonlinear

transformations of the original stimuli. The dashed lines sketch the vector K. **C.** 1D projections of the plots in **B** along K. **D.** Model interpretability. For the linear model (left), K is defined in the space of the stimuli, so looking at its structure is sufficient to obtain insight on the model computations (e.g. preferred polarity, surround strength). For the nonlinear model (right), K depends on the nonlinear transformations preceding it and this impairs interpretability.

## 3. Success and limits of normative approaches

The retina is the first stage of processing of the visual system. Therefore, one of the main purposes of the retina is to transmit visual information to the rest of the brain. For this, it needs to compress the information received by $10^8$ photoreceptors into spiking activity sent through the optic nerve by $10^6$ ganglion cells. The efficient coding approach (Attneave 1954; Barlow 1961) aims to understand retinal function as the optimal way to achieve information compression, independently from specific features. Classically in this approach an objective function is maximized, under some defined constraints, and the results predict some aspects of ganglion cell function (reviewed in (Simoncelli and Olshausen 2001; Karamanlis et al. 2022; Manookin and Rieke 2023). Any formulation of efficient coding comes with four key components: 1) an objective function to be maximized (e.g. mutual information between stimulus and response); 2) some constraints that must be respected, or a cost to be minimized (e.g. a metabolic cost associated with every spike emitted) ; 3) a statistical description of the natural input neurons are exposed to; 4) an assumption on the functional form through which the recorded neurons encode their inputs (Figure 2 ; see Młynarski 2025).

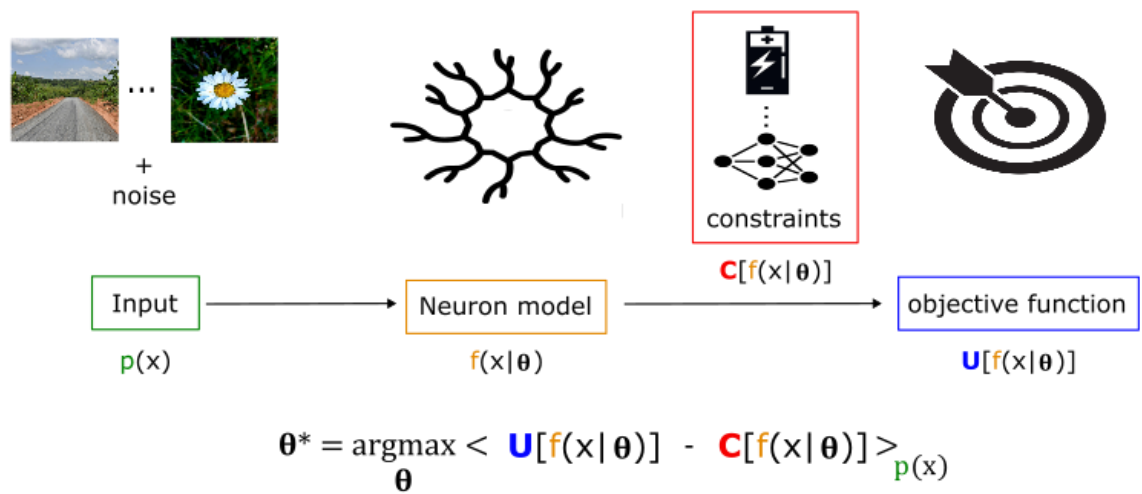

**Figure 2: Normative approaches scheme** A normative theory of sensory processing requires specifying four key components: (1) the input distribution p(x), describing the statistics of natural stimuli; x corresponds to a stimulus (2) an encoding model f(x|θ), describing how stimuli are transformed by the system into spikes; $\theta$ corresponds to the model parameters (3) an objective function U[f(x|θ)], quantifying the system performance on a given task (e.g. information transmission); and (4) a set of constraints C[f(x|θ)], representing biological or energetic limitations. The goal is to find the parameters $\theta^*$ that maximize the expected objective function minus the cost, across the input distribution.

Since natural inputs are redundant (i.e. highly correlated is space and time), for a low noise level, efficient coding predicts that a ganglion cell should reduce the correlation of natural inputs in its outputs with minimal loss of information. Despite this, efficient coding should not be reduced to decorrelation. Early theoretical work has argued for example that complete stimulus decorrelation in the retina is not always optimal due to transduction noise (Atick and Redlich 1992). Enforcing this normative constraint, leads to the prediction of the center-surround receptive field structure observed experimentally across noise levels (Kuffler 1953; Barlow et al. 1957; Enroth-Cugell and Robson 1966; Rodieck 1967; Atick and Redlich 1992).

Further imposing that information transmission should happen while emitting as few spikes as possible, successfully explains a number of basic aspects of retinal processing like why ganglion cells split in ON and OFF types (Gjorgjieva et al. 2014), why OFF cells are often smaller and more numerous than ON cells (Karklin and Simoncelli 2011), why ON and OFF cells mosaics are anti-aligned, i.e. the receptive field centers between mosaics are more distant than expected by chance (Roy et al. 2021), how their alignment depends on input noise (Jun et al. 2021) and their observed distribution with retinal eccentricity (Ocko et al. 2018). Similar approaches also partially account for ganglion cells further specialization in different functional types beyond ON and OFF (Kastner and Baccus 2011; Ocko et al. 2018; Jun et al. 2022; Młynarski and Hermundstad 2021). They successfully explain the division in transient vs sustained cells and some observed properties of temporal adaptation but without giving a full account of the diversity characterized in the retina.

To summarize, the efficient coding approach is a useful and successful theory that predicts several observed computations (reviewed in (Turner et al. 2019; Karamanlis et al. 2022). However, it comes with several limitations.

### 3.1. Beyond quasi-linear models

The vast majority of efficient coding studies assume that retinal ganglion cells process visual stimuli in a linear or quasi-linear manner (i.e., LN model,(Karklin and Simoncelli 2011; Gjorgjieva et al. 2014; Roy et al. 2021)). While this approximation is reasonable for some simple artificial stimuli (Wienbar and Schwartz 2018), it becomes overly simplistic when applied to natural stimuli. As we have reported above, retinal processing of natural stimuli is highly nonlinear, at least for specific ganglion cell types (Turner and Rieke 2016; Karamanlis and Gollisch 2021). Importantly, this nonlinearity is not limited to "lower" species like cat, rat, mouse, guinea pig, or rabbit, but is also evident in the primate retina (Heitman et al. 2016; Turner and Rieke 2016; Gogliettino et al. 2024).

Neglecting these nonlinearities in efficient coding approaches can significantly limit their predictive power for several reasons. First, nonlinearities play a crucial role in the amount of information that the system can transmit (Pitkow and Meister 2012). Second, different types of ganglion cells can exhibit similar responses when probed with simple stimuli while more natural ones would reveal distinct nonlinear computations (Gollisch and Meister 2010; Kastner et al. 2015; Trapani et al. 2023). This may explain why efficient coding approaches have struggled to account for the extensive diversity of ganglion cell functional types observed in experimental data— at least 32 in the mouse (Baden et al. 2016) and 23–32 in the macaque (Kling et al. 2024): when reduced to quasilinear models, distinct types can become very similar.

To accurately relate the characteristics of retinal processing observed during natural vision to normative principles, efficient coding frameworks must be refined to incorporate the diversity and nonlinear properties of retinal ganglion cell populations. However, models that include these non-linearities can have a large number of parameters, which can be a challenge for normative theories. Optimizing the value of a function with many variables can have many solutions. If we try to derive the model parameter values from the optimization of an objective function, we might obtain several solutions, and the real solution may not even be among them (Młynarski et al. 2021). More parameters will require more constraints to tease apart which solution is closest to reality.

### 3.2. What is a natural stimulus and the limitations of conditioning on a stimulus distribution

A description of stimulus statistics is a critical component of efficient coding approaches. Most studies seek to derive predictions under the assumption of natural stimulus statistics, as these are the most ethologically relevant. However, defining what constitutes a natural stimulus is not straightforward.

A common approach is to consider as natural stimuli the visual scenes captured with a camera in everyday life (van Hateren and van der Schaaf 1998). Some statistics, like 1/f spectrum, seem to be universally shared by most natural scenes (Field 1987; Simoncelli and Olshausen 2001). However, the hypothesis that all natural stimuli have the same statistics has limitations.

Even the same scene can appear differently depending on the observer. Differences in body size, eye optics, and typical posture may result in distinct visual inputs for different animals, even within the same environment. These static body constraints can reshape retinal selectivity.

For example, eye size seems to exert a key influence on the speed tuning of direction-selective ganglion cells (DSGCs). The preferred speed of DSGCs varies systematically with eye size so that this tuning remains consistent across mouse and rabbit when expressed in terms of stimulus motion across the visual field rather than retinal velocity (Ding et al. 2016). In addition, the optical properties of the eye, which vary greatly across species due to evolutionary and behavioral constraints, act as a spatial (Goethals et al. 2025) and partially as a temporal (Fitzpatrick et al. 2024) filter on incoming visual scenes and affect visual acuity.

Species-specific body posture also affects visual statistics. For example, many quadrupeds have their visual field split by a horizon line, with the upper part dominated by sky and the lower by land. This ecological constraint contributes to the organization of center-surround interactions across the retina (Gupta et al. 2023).

Beyond static features, dynamic constraints such as movement also reshape retinal input. Visual input is modulated in time by eye movements, like saccades and fixational drifts, which introduce additional correlations and discontinuities in the visual stream (Van Der Linde et al. 2009; Rucci and Victor 2015; Turner et al. 2019). Moreover, the way animals coordinate eye and head movements to explore their environment further influences the sequence of visual inputs reaching the retina

(Skyberg and Niell 2024). These factors can fundamentally alter the statistical structure of the stimuli that retinal circuits process, especially over short timescales.

Other whole-body movements, such as walking, further influence visual input. The regular pattern of motion during locomotion modulates the structure and rhythm of sensory inflow (Hayhoe and Ballard 2005), adding another layer of species-specificity to the statistics experienced by the retina. Finally, the optimal way of encoding natural stimuli is also shaped by an animal's visual tasks. Different tasks impose distinct perceptual demands, which affect both the frequency and importance of specific stimuli (Muller et al. 2023).

These observations highlight the need for species-specific datasets of natural statistics (Qiu et al. 2021), particularly when considering animals such as birds or aquatic species, whose environments differ fundamentally from those of terrestrial mammals. They also suggest that further progress can be made by incorporating species-specific constraints and task-dependent visual goals in normative approaches.

## 4. Towards task reductionism

### 4.1. Visually-driven behaviors and cell types

Animals perform a wide range of visually guided behaviors, each of which can modulate both the structure and relevance of visual input. For example, during prey pursuit, self-motion dramatically alters the visual statistics compared to when the animal is stationary and scanning for predators. Behavior not only reshapes sensory input but also alters the relative importance of different visual features. For instance, detecting an approaching predator may require heightened sensitivity to rare but critical stimuli. In such cases, allocating more coding resources to these low-probability events may be beneficial, even if it deviates from strict statistical efficiency. This suggests a limitation to efficient coding frameworks, which typically optimize for information transmission without incorporating behavioral relevance. Prior work outside the retina field has already suggested the importance of integrating behavioral demands into coding principles (Machens et al. 2005).

This issue was known even before efficient coding and led to the formulation of an alternative theory: the idea that ganglion cell types extract features that are directly useful for specific visually driven behaviors (Lettvin et al. 1959; Riccitelli et al. 2025). In their classical work, Lettvin et al (1959) identified 4 different visually-driven behaviors in the frog, and found 4 cell types that each extracted a feature perfectly suited for these 4 tasks. For example, a specific class of ganglion cell would respond optimally to the presentation of a dark spot similar to a bug, and would elicit a striking behavioral response of the frog (Barlow 1953). This direct 1-to1 connection between cell type and behavior was deeply influential, and suggested that there could be "labeled lines", i.e. separate channels of visual information where separate cell types would extract distinct features, that would serve different visual behaviors.
Several other studies also suggested a link between one cell type and one visually-driven behavior, or a perception (Barlow 1972). For example, ON DS cells seem to be causally involved in the optokinetic reflex and parasol cells might be key to detect motion (Merigan and Maunsell 1990; Merigan et al. 1991).

However, there are cases in which more than one cell type may be involved in a single behavioral response. For example, both W3 and tOff alpha ganglion cells seem to be involved in the escape response to a looming stimulus. Although W3 might be more involved in the detection, and tOFFaplha in the speed encoding (Kerschensteiner 2022), both are informative about the looming stimulus, and the two can be perturbed by removing VGlut3+ amacrine cells, which does impact the behavioural response to looming (Münch et al. 2009; Kim et al. 2020; Kerschensteiner 2022). Perturbing tOff alpha alone (Wang et al. 2021) also alter the behavioral response to looming. Several cell types might also be involved in the ability of mice to catch a cricket in an arena (Johnson et al. 2021; Skyberg and Niell 2024).

Conversely, the same cell type can be used for different tasks in different visual contexts. tOff alpha cells, for example, are not only involved in detecting looming threat, but are also specifically sensitive to image recurrence across saccadic eye movements (Krishnamoorthy et al. 2017). Therefore, the idea of strict "labeled lines" (Adrian 1928; Barlow 1972) is incomplete. While some ganglion cells may be specialized for specific tasks, others appear to support shared or multiplexed functions, depending on behavioral context. This complexity is particularly relevant in species capable of many visual behaviors—such as humans—where it is unlikely that each behavior corresponds to a unique ganglion cell type.

These findings suggest that both the efficient coding framework and the "labeled lines" framework are incomplete. Efficient coding does not capture behavioral relevance, while labeled lines fail to account for shared or context-dependent roles of cell types.

### 4.2. An entangled relation between cell types and visual behaviors

We propose a more general framework, that includes both the labeled lines and the efficient coding hypotheses as special cases. The original labeled lines hypothesis (Adrian 1928; Lettvin et al. 1959) suggested that each cell type was somewhat dedicated to a specific visual task. This assumption is too rigid to account for the diversity of tasks that animals can achieve. In contrast, we suggest that each natural visual behavior recruits a subset of ganglion cell types, and that subsets associated with different behaviors can partially overlap. This many-to-many mapping gives rise to an entangled relationship between cell types and behaviors: a single behavior may rely on multiple cell types (divergence, Figure 3), and a single cell type may contribute to multiple behaviors (convergence, Figure 3). This view is supported by recent work in *Drosophila* where overlapping subsets of visual projecting neurons drive different types of social interactions (Cowley et al. 2024). In Cowley et al (see their fig 5a) each visual feature is represented by several cell types, and each social behavior is driven by a subset of cell types.

In our framework, a single cell type can participate in either one, few, or many visually-driven behaviors. If a cell type is involved in only one behavior, and if the behavior only relies on this cell type, the labeled lines hypothesis provides a plausible explanatory framework. However, this hypothesis needs to be expanded: there are cases where a single visual task will be achieved by relying on the convergence of the activity of several cell types.

At the other extreme, if a cell type, or a group of cell types, is broadly used across many tasks, optimizing its coding strategy for each of these many tasks may be well approximated by optimizing it for information transmission about the input itself. In this case, the efficient coding framework might provide a good explanation. This may apply to primate parasol and midget cells, which are involved in many visual behaviors. Previous work has shown that their spatial and temporal selectivity can be captured, at least in part, by models that optimize information transmission about natural inputs (Ocko et al. 2018; Jun et al. 2021). From this perspective, efficient coding succeeds not because it explicitly captures task constraints, nor because optimizing information is the primary objective of sensory neurons, but because it approximates the requirements of many tasks through a single information-theoretic objective. This simplification enables the use of a single objective function and has proven powerful in explaining several functional properties of some retinal ganglion cell types.

However, its limitations may become apparent when dealing with intermediate cases—where cell types are involved in a small subset of behaviors, and their coding properties are shaped by a more specific set of behavioral demands. In such cases, the cell's coding may reflect the converging of specific constraints from the few tasks it supports. Computational studies have shown that mixed selectivity—tuning to combinations of features—can emerge when neural networks are jointly optimized for multiple tasks (Yang et al. 2019; Johnston and Fusi 2023; Feather and Chung 2023). This divergence from one cell type to several tasks may be essential to explain the tuning properties of some retinal ganglion cell types, whose responses cannot be understood from a single behavioral role or information theoretical considerations.

For this reason, several recent studies (P. R. Parker et al. 2022; P. R. L. Parker et al. 2022; Skyberg and Niell 2024; Riccitelli et al. 2025) have argued for a return to ecologically grounded tasks, especially in cases where behaviorally relevant computations can be defined and tested. Under this perspective, cell types may be better understood by asking which tasks they support and how these tasks shape encoding.

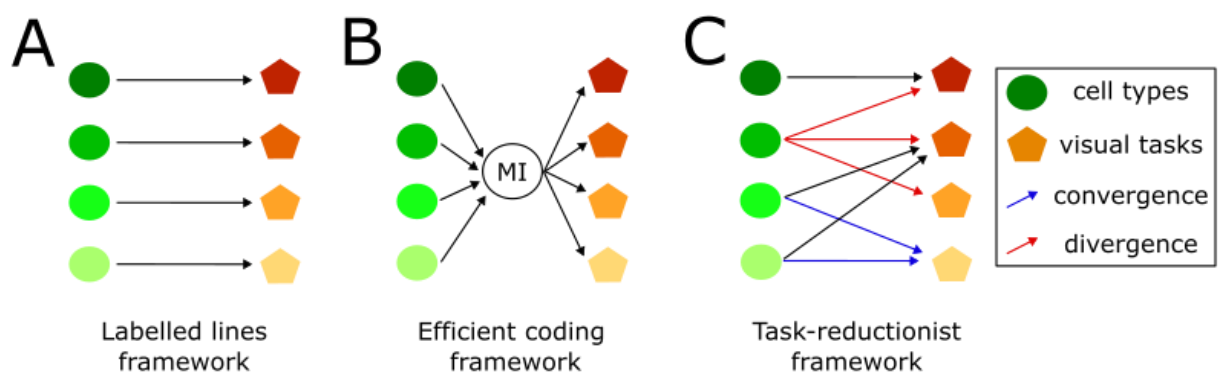

**Figure 3: A novel normative framework: Task reductionism** Sketches of the different normative frameworks proposed to study retinal processing. **A.** Labelled lines: each cell type is devoted to a different task **B.** Efficient coding: All the cell types are optimized solely for information transmission (e.g. Mutual Information between stimuli and responses, MI) which then allows the rest of the brain to support every task **C.** Task reductionism: A task can be supported by several cell types (convergence) and a single cell type can be involved in many tasks (divergence).

### 4.3. Understanding function through the lens of natural behaviors

This framework suggests a "divide and conquer" strategy towards understanding how the different ganglion cell types process natural scenes. Rather than decomposing the natural scenes into simpler components, a possible way forward is to divide the problem in the different visual tasks performed by the animal. A specific type might be better understood by replacing efficient coding by objective functions that are more precisely connected to specific tasks. Instead of maximizing the amount of information transmitted, the objective function to optimize should be related to a visually driven behavior that has a clear ethological relevance. For example, one objective function could be the ability to locate where the prey is during a prey catching task. Ganglion cell types could then be evaluated by their ability to contribute to this task, both with normative and encoding approaches.

While studying ganglion cell function through the lens of visual task has been done in the past (Lettvin et al. 1959; Riccitelli et al. 2025), the field is now ripe for a renewal of this kind of approach for several reasons:

1) several visually driven behaviors have been or are currently being extensively characterized. Focusing on ethologically relevant behaviors make sense, because only these ones are likely to be optimized throughout evolution. For example, several studies have characterized natural visually-driven behaviors in the mouse (Clark et al. 2006; Yilmaz and Meister 2013; Hoy et al. 2016; Pakan et al. 2018; Boone et al. 2021; Ebbesen and Froemke 2021; P. R. Parker et al. 2022).

2) several tools (body and eye tracking systems) now allow to reconstruct the retinal input during visually driven behaviors (Holmgren et al. 2021; Ressmeyer et al. 2025).

3) it is now becoming possible to inactivate specific types and understand which types are necessary for which tasks (Wang et al. 2021; Johnson et al. 2021).

4) machine learning tools allow to model better how ganglion cells process complex visual scenes like the ones that are projected on the retina during realistic visual tasks (Karamanlis and Gollisch 2021; Goldin et al. 2022; Maheswaranathan et al. 2023; Höfling et al. 2024; Vystrčilová et al. 2024).

Replacing the mutual information with more task-driven objective functions suggests several ways to move forward for both normative and encoding approaches.

### 4.4. Optimizing information… relevant for a visual task

In normative approaches, we may need to go beyond the classical objective of information maximization and start to ask "information about what ?". If specific subsets of cell types are optimized for specific tasks, they might be optimized to transmit information about specific features in the visual context corresponding to these tasks. For example, some cell types might be optimized to encode information about the position of the prey in the cricket hunting task (Hoy et al. 2016).

A first step in this direction has been taken by asking if neurons were optimally encoding the future stimulus (Palmer et al. 2015; Chalk et al. 2018; Manookin and Rieke 2023), especially for stimuli moving randomly. This objective function makes sense for any task that requires tracking moving objects. Different cell types optimally encode the future stimulus, while others do not, and this for specific types of motion (Liu et al. 2021). The advantage of predictive information is that it still remains quite agnostic about the feature being encoded. However, future studies may have to look for a more direct connection to specific tasks.

Defining an objective function in terms of a specific task also changes what is considered signal and what it is considered noise. Classically, noise corresponds to the random fluctuations in neural responses that cannot be controlled even when repeating the same exact stimulus (trial-to-trial variability). However, when trying to find the optimal way to solve a task, some features of the stimulus are irrelevant as they are not informative for the task to solve. Therefore, from a theoretical point of view, they should be considered noise. In this view, noise corresponds more generally to all the stimulus features that are not useful to solve the task at hand. By changing the definition of what is considered signal and what noise, optimal solutions will change. For example, during daylight, the classical signal to noise ratio (SNR) of retinal responses is typically considered high, as trial to trial fluctuations are relatively small compared to the mean response. However, if every irrelevant aspect of the visual scene to a given task is considered as noise, the effective signal to noise ratio can be much lower.

The implications of this change of SNR have already been shown in the fly, where (Sinha et al. 2021) showed that an optimal motion encoder has a very different form if noise is defined as "all the sources of variability in the natural scene that are irrelevant to global motion". In this study, the authors aimed at extracting the global motion from the reconstructed retinal input. The input to a photoreceptor can vary for a lot of reasons: if there is a global motion, but also if there is a local change in luminance, or a global change in lighting. The authors pooled together all the sources of variability that are not due to global motion, and consider all of them as noise. As a result, the signal to noise ratio became much lower than if noise was more classically defined as the trial-to-trial variability across stimulus repetitions. This change of signal to noise ratio means that, from a theoretical point of view, the system is viewed as being in a high noise regime, where optimal solutions can be qualitatively different from a low noise regime. This highlights how redefining noise in task-relevant terms can place the system in a high-noise regime even under conditions, like daylight, where classical definitions would not.

This shift in perspective has important implications for theory. Several studies have shown that, in models of retinal processing, qualitatively different solutions emerge when trying to solve the same problem in a high noise or low noise regime (Atick and

Redlich 1992; Tkacik et al. 2010). This change of point of view, where noise corresponds to "everything that is irrelevant", and thus is typically much higher, might give novel predictions about how the retina should be organized.

Finally, this change of perspective on the noise can also give mechanistic insights about retinal organization. Within the circuit generating direction selectivity, it has been shown (Chen et al. 2016) that inhibitory synapses between On starburst amacrine cells are not necessary for direction selectivity of the On pathway when the moving object is a bar on a uniform background. However, these synapses become necessary to maintain direction selectivity when the background is noisy. This shows that part of the retinal circuit may not be designed for feature selectivity per se, but rather to make this feature selectivity robust to other sources of variability in the stimulus – in this case, to a white noise added on the background of a moving bar. Other aspects of the retinal circuit may also be best understood when trying to maximize information about a feature – such as motion direction – while trying to be robust and insensitive to other aspects of the stimulus, such as background noise.

### 4.5. Visual tasks to understand encoding models

Focusing on how a cell type is useful for a specific visual task may also improve the functional interpretability of encoding models and might help with the issue of their increasing complexity. As models become increasingly high-dimensional, understanding what a cell "does" in general becomes more difficult. This is particularly true when models are trained on natural stimuli, which span a wide range of visual contexts and features.

While in some cases, controlled natural image manipulations can reveal simplified models that qualitatively explain ganglion cell behavior (Turner and Rieke 2016; Goldin et al. 2022), there are other cases where no compact model can give an account of how a ganglion cell responds to natural scenes. In such cases, the result of modeling may be a complex function that resists low-dimensional interpretation. When this happens, a possible strategy is to shift the question: instead of asking how a cell responds to arbitrary stimuli, we could ask: how useful is the cell's response for solving a specific visual task?

A possible strategy is to look at what information is carried by a certain ganglion cell type when stimulated by visual inputs corresponding to a specific visual task. A same ganglion cell type can support a given function when having to detect a predator, and another one when trying to catch a prey. It has already been shown how a single ganglion cell type can switch from encoding one feature to another when the visual context changes (Geffen et al. 2007; Deny et al. 2017; Krishnamoorthy et al. 2017). These cells multiplex the information about different features depending on the visual context.

A possible way forward to demultiplex this information could thus be to analyze how informative / useful a ganglion cell type can be to solve a given task. This might make the interpretation of what a ganglion "does" simpler. First, focusing on a visual task reduces the dimensionality of possible visual inputs: stimuli encountered during a specific task like cricket hunting are more similar to each other than those drawn from an unconstrained natural movie dataset. Reduced to this smaller input space, encoding models might be approximated by simpler descriptions.

Second, having a specific task to achieve provides an objective function and allows to quantify how useful a certain type is to achieve the task compared to other types. This can be done by quantifying how well the task can be achieved by linearly decoding the types activity (Oesterle et al. 2025), or by quantifying the mutual information between the task relevant features and the ganglion cell activity.

Finally, this task-based framework could also help for population models, e.g. when the noise correlations between cells are included. Having an objective function derived from a visual task can help understanding the role of the different components of these population models. It might be that noise correlations are useful for some tasks, and not for others (Mahuas et al. 2024).

Overall, restricting the analysis to specific visual tasks may not only simplify the stimulus space but also provide a tractable framework for interpreting complex models. This approach may be particularly useful for cell types whose general function cannot be easily described in low-dimensional terms.

### 4.6. Cell types contributing to multiple tasks

While this reductionist strategy of defining a visual task and then analyzing the contribution of each cell type in order to understand cell type computations is appealing and powerful, it is important to acknowledge its limitations. Some aspects of retinal organization, and some ganglion cell types, may only be understood as the result of joint optimization across the multiple tasks in which they are involved, due to the divergence between cell types and tasks. Several works showed that retinal circuits can be recycled for different purposes in different contexts. For example, the same circuit formed by AII amacrine cell and cone bipolar cell that relays rod bipolar cell signal at scotopic light level, is used at photopic light level to shape the selectivity of tOFF alpha cells by making them selective to approaching motion (Münch et al. 2009). Similarly, a single cell type can encode different features in different visual contexts (Deny et al. 2017). Computational studies have shown that specific types of selectivity only emerge when optimizing the network model for multiple tasks at the same time (Yang et al. 2019; Feather and Chung 2023; Johnston and Fusi 2023). This is the case of the so-called "mixed selectivity", where a neuron responds to multiple features.

Another intriguing type of selectivity that emerges optimizing a network model for multiple tasks is the selectivity to "abstract" features (Johnston and Fusi 2023). Some neurons become selective to features that cannot be related to a specific task. The advantage of these "abstract representations" is that they allow the network to perform multiple tasks at once, but also to generalize better: new tasks can be learned with a few trials, and it seems that these abstract representations are keys to achieve this (Johnston and Fusi 2023). It is thus possible that some ganglion cell types, being used for several different visual tasks, will be selective to features that cannot be directly connected to a single task. These abstract representations are not necessarily "platonic" representations (Huh et al. 2024), but just a byproduct of having to reuse the same neurons and circuits for multiple tasks.

A possible speculation is thus that some ganglion cell types will only be fully understood as the product of a joint optimization over several tasks to which they participate. This might be the case of OS cells, as extracting orientation is probably useful for various tasks. Their selectivity may reflect an optimization that balances performance across these tasks. Understanding these cell types may require moving beyond task-specific models and considering the constraints imposed by shared functionality across behaviors.

In summary, the models employed to predict ganglion cell responses to natural scenes have been increasingly complex. DNNs have a large number of parameters, which makes them difficult to interpret, both for mechanistic insight and to understand their function.
Our hypothesis of an entangled relation between cell types and tasks, with both convergence and divergence motifs, suggests a way forward.
Focusing on the convergence motifs, the model complexity might be overturned by focusing on a single visual task and estimating the usefulness of each type to perform this task. This is becoming now possible for several reasons. Ethologically relevant visual tasks are now being characterized in multiple species, and it becomes possible to reconstruct the visual input to the retina during visual tasks with a greater precision than before. Machine learning models offer good approximations of retinal processing of these complex visual inputs. New tools allow to isolate which cell types are causally involved in a visual task. Together, these advances will allow to quantify the role of each ganglion cell type in a given task with unprecedented precision.

Conversely, the divergence motifs may allow to understand more abstract feature selectivity (i.e. that cannot be connected to a single task). It is likely that some ganglion cell types will only be fully understood as jointly optimized for several visual tasks.

## 5. Integrating new constraints in normative and encoding approaches

As we have described above, a major issue to make progress both for encoding models, and for normative approaches, is model complexity. For encoding models, this leads to an increase in the number of parameters that make generalization, as well as interpretability, more challenging. For normative approaches, it makes it difficult to predict the non-linear computations performed by ganglion cells, because they can only be described by models with a large number of parameters. In both cases, the large number of parameters leads a space of possible solutions that is too large. A promising strategy is to constrain this space and reduce this degeneracy by integrating constraints from biology. Here we will review three types of data that can be used as constraints: connectomics, evolution, and physiology experiments.

### 5.1. Connectomics to constrain normative approaches

Thanks to the tremendous progress in characterizing the anatomy of neural circuits (Helmstaedter et al. 2013; Sigulinsky et al. 2024; Dorkenwald et al. 2024; Bae et al. 2025), a possible way forward for normative approaches is to integrate constraints that are much more detailed and closer to biology: instead of trying to recover retinal computations ex nihilo, it becomes possible to take as constraints some of the known features of the retinal circuit, and see if, together with an optimization principle (maximizing mutual information, or task performance), they can predict some aspects of the retinal code.

A long-standing hypothesis is that the simplest way to describe the function of a neural network might be to describe the connections between the neurons, rather than the activity patterns that it produces (Von Neumann, 1949). With connectomics, it might soon become possible to know the full wiring diagram of all the circuit shaping the response of ganglion cell types, together with their morphology. By itself, this connectome is extremely informative, but it is unlikely that this will be enough to build a model of the computations performed by ganglion cells with a bottom-up approach, simply because too many parameters (e.g. synaptic weights, ion channels) will have unknown values. However, we can assume that the missing parameters have been optimized by a principle such as efficient coding, or task performance, and obtain a model closer to biology this way.

This approach has already been successfully tested in the fly, where connectomics efforts are more advanced. Lappalainen et al took the connectome of the fly visual system as a constraint for their model, and then learned the remaining parameters by optimizing the performance of the model at doing motion inference (Lappalainen et al. 2024). They could predict many properties of the cell types of the fly visual system. Outside of neuroscience, a recent study has shown that it is possible to maximize information transmission in a detailed model of a biochemical network and obtain prediction and explanation about how this network is organized (Sokolowski et al. 2025). This shows how combining an optimization principle with the constraints of the connectome can allow to predict the function of several cell types.

With the on-going efforts in mapping the full connectome of the mammalian retina (Helmstaedter et al. 2013; Sigulinsky et al. 2024; Dorkenwald et al. 2024; Schlegel et al. 2024), it might be soon possible to perform similar attempts in the retina, using either efficient coding or task performance as an objective function. This would mean to include detailed connectivity constraints and optimize the unknown parameters (synaptic weights, and possibly ion channel expression levels) thanks to an optimization principle. If the purpose is to have a detailed biophysical model, gene expression maps might also be used as additional constraints to define which ion channels are expressed in which cell types. Inactivation of well-defined cell types could also be included in this attempt to better map specific units of the model to well-defined cell types in the retinal circuit (Cowley et al. 2024).

This approach also gives a different view on normative approaches: instead of using an objective function to recreate a biological system ex nihilo, here the optimization principle can be seen as a way to stir a given biological system with wiring constraints towards an objective. Such an approach might be closer to how optimization worked across evolution.

## 5.2. Integrating constraints from evolution in normative approaches

In sensory systems, it is often the case that multiple distinct circuit architectures could, in principle, implement the same computation. This functional degeneracy poses a challenge for normative approaches, which typically aim to characterize a system by optimizing an objective function under the assumption of functional optimality (Figure 2). When such optimization is performed over a network with a high-dimensional parameter space, it frequently yields not a single unique solution, but a large set of near-optimal solutions (Sokolowski et al. 2025). Even when a global optimum exists, many parameter configurations may produce behavior or responses that are nearly indistinguishable in terms of objective function value—yet only one of them corresponds to the biological implementation observed.
A plausible explanation is that evolutionary processes do not search the parameter space from scratch; instead, they operate through modification of existing structures. Among many functionally similar solutions, evolution tends to favor those that are incrementally reachable from ancestral architectures. This phenomenon, known as "historical contingency" in evolutionary biology, suggests that the observed solution may not be the most optimal in an abstract sense, but the most accessible given the organism's phylogenetic history. This contingency may also explain some aspects of retinal organization.

A recent study in insect vision (Buerkle and Palmer 2020) exemplifies this principle. The authors show that although there are multiple viable implementations of trichromatic or tetrachromatic vision in insects, the specific architecture adopted in a given species can be explained by incorporating constraints from phylogenetic lineage. Similarly, in vertebrates, color processing varies widely across species. While efficient coding theories often attribute this diversity to adaptations to species-specific visual environments, it is likely that evolutionary history also constrains the range of implementable solutions. For instance, the transition from dichromacy to trichromacy has occurred independently in multiple lineages and sometimes long after the establishment of dichromatic systems. Incorporating such temporal order constraints may help explain the organization of color processing circuits in specific species.

More broadly, the efficient coding hypothesis assumes that the retina has been optimized for information transmission over evolutionary time. While most approaches focus solely on the end point of this process —i.e., the current circuit architecture — in some cases it may be equally important to consider the trajectory by which the system evolved. Integrating phylogenetic constraints into normative models could therefore yield more accurate and biologically plausible explanations of sensory circuit organization.

## 5.3. Connectomics to constrain encoding models

Structural data from connectomics also offers a promising avenue for introducing biologically meaningful constraints into DNNs. This is particularly relevant for the early visual system, where detailed circuit reconstructions are increasingly available. Recent work in the fly (Cowley et al. 2024; Lappalainen et al. 2024; Seung 2024; Christenson et al. 2024) and in the mouse visual cortex (Wang et al. 2023; Bae et al. 2025; Ding et al. 2025) has shown that integrating connectomic data into network models can lead to more interpretable architectures as it allows mapping specific components of the model to cell types in the circuit (Christenson et al. 2024). Methods allowing to perturb or eliminate specific cell types (Johnson et al. 2021) or connections (Chen et al. 2016) may also provide constraints for this type of approach. This has the potential to solve at least partially the aforementioned problem of mechanistic interpretation of complex models: by assigning each model component to specific cell types, this makes the mechanistic interpretation possible.

A first attempt in this direction in the retina was carried out by (Schwartz et al. 2012): by constraining a network model with the reconstructed connectivity matrix between bipolar and ganglion cells of a given type, the model could predict how a ganglion would respond to textures of various orientations. More detailed information on the connectivity will allow to make progress in this direction, moving from descriptive to mechanistic models of retinal computations.

An additional benefit of such structural constraints is the potential to improve generalization across stimulus domains, a challenge faced by many existing models. By aligning the model architecture more closely with biological circuitry, such constraints may reduce the risk of overfitting and enhance the robustness of model predictions.

However, one fundamental reason for limited generalization in current network models is the absence of adaptive mechanisms that are intrinsic to biological circuits. In real neural systems, changes in input statistics lead to circuit-level adaptations mediated by processes such as dynamical synapses (Beaudoin et al. 2008; Ebert et al. 2024) or adaptive intrinsic properties of single neurons (Brown and Masland 2001; Kim and Rieke 2003; Manookin and Demb 2006). These dynamic properties are not fully specified by static connectomic data and may therefore limit the ability of connectome-constrained models to generalize to new stimulus regimes.

Despite these limitations, leveraging connectomics remains a promising strategy for improving both the interpretability and generalizability of encoding models. By reducing the degrees of freedom in the model space and anchoring computations to anatomical structure, such constraints may ultimately lead to more biologically plausible and robust models of sensory processing.

### 5.4. Natural stimuli manipulation to constrain models of retinal processing

Another line of research to constrain models of retinal processing is to extract biologically some qualitative functional insights by applying structured manipulations to natural stimuli. As mentioned above, when training models with many parameters on the responses to complex stimuli, problems such as overfitting and limited functional interpretability arise. Manipulating natural scenes in controlled ways can expose aspects of retinal computations that help restrict the space of valid models, and help functional interpretation.

The standard method to evaluate models, including deep neural networks, is to test their predictive accuracy on held-out data. The implicit assumption is that high correlation between predicted and observed responses indicates that the model representation that leads to successful predictions should be aligned with the neural representation. This assumption is not always verified. Observed correlation is never one. Even when it is high for a certain stimulus statistic it often drops when the model is tested out of distribution (here a recent example in the retina (Vystrčilová et al. 2024).

A more informative approach is to directly characterize the function being modeled. Let F be the tuning function that maps stimuli to neural responses, either for individual neurons or populations. For natural stimuli, F is both high-dimensional and highly nonlinear, already at the level of the retina. Instead of only fitting F with a model, one way to characterize F experimentally is to understand better what its gradient looks like. The gradient of F indicates the direction of change of the stimulus that will increase the most the response. Directions orthogonal to the gradient define invariances—stimulus changes that leave the response unaffected. Probing directly invariance and selectivity (i.e. directions of maximal changes) have been promising avenues of research in sensory processing, and in the retina.

When trying to extract information about a feature, retinal neurons leave out visual information about other aspects of the stimulus. As a result, they are invariant to some changes in the stimulus. Probing this invariance experimentally is a way to gain insights on how a ganglion cell processes visual information. Trying to find two stimuli that evoke the same response in a cell is an idea that was originally developed to study color processing in visual psychophysics, by finding different color spectra that would evoke the same percept, called color metamers. Metamers have then been used to understand how higher areas of the visual system processes visual information (Freeman and Simoncelli 2011). In the retina, (Turner and Rieke 2016) used it to test if natural images are integrated linearly or non-linearly in the receptive field center of parasol ganglion cells. They presented a natural image inside the receptive field center, and then a "linear equivalent" stimulus, which should evoke the same response if the cell integrates the visual stimulus linearly. This allowed them to demonstrate non-linear spatial integration in Off parasol cells, and linear integration in On parasol cells. In further studies (Yu et al. 2022), using the same technique, they could show how this integration could be modulated by surround stimulation, and that the asymmetry was at least partially due to the non-linear processing of photoreceptors. The interest of this technique is to avoid using artificial stimuli to probe spatial integration (since they could give different results than natural ones (Yu et al. 2022), while still being able to test direct hypotheses through relatively simple stimulus manipulation. It replaced the simplicity of artificial stimuli with simple manipulation of natural stimuli. Their results put constraints on the models that want to predict how parasol cells process natural stimuli.

Another way to constrain models is to use the complementary approach, i.e. find the change in the stimulus that will evoke the strongest change in the response. A recent line of work has successfully identified the maximally exciting inputs of sensory neurons (Walker et al. 2019; Ponce et al. 2019; Bashivan et al. 2019). Specifically, after training a deep neural network to approximate F, one can compute the model's gradient and follow it iteratively from a random starting point in stimulus space until convergence. The resulting stimulus, known as the Maximally Exciting Input (MEI), is predicted to evoke the strongest response in the neuron and the prediction is verified experimentally. This approach has been used in the retina to link suppressed-by-contrast RGCs to a form of green-UV color opponency (Höfling et al. 2024) and in the visual cortex to reveal complex feature selectivity across species and brain areas (Ding et al. 2023; Willeke et al. 2023; Fu et al. 2024). Maximally Discriminative Stimuli have extended this approach to find stimuli that strongly drive specific cell types while leaving the others unaffected (Burg et al. 2024).

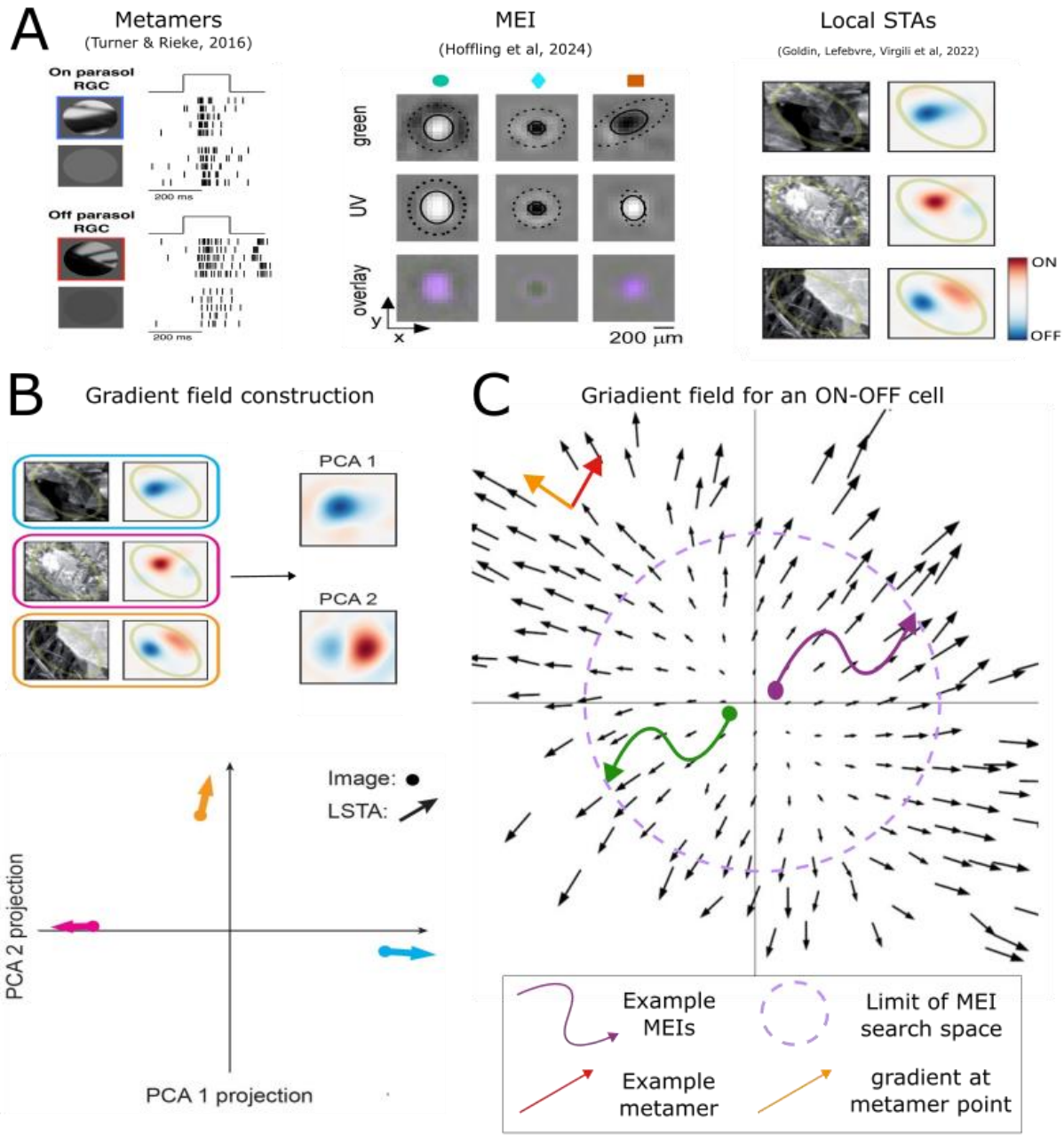

**Figure 4: Probing the functional selectivity and invariance of retinal ganglion cells using stimulus manipulations.**
**A.** Different strategies to characterize the tuning of retinal ganglion cells using controlled manipulations of natural stimuli. Left: *Metamers* (Turner & Rieke, 2016) test for response invariance by presenting natural images and their linear equivalents. Center: *Maximally Exciting Inputs (MEIs)* are computed by gradient ascent on a modelled cell to identify stimuli that maximize the response (Höfling et al., 2024). Right: *Local STAs* (LSTAs) are data-driven estimations of the gradient calculated around several natural images (Goldin et al., 2022). They capture the local structure of the cell's sensitivity in different stimulus contexts. **B.** Construction of a gradient field from LSTAs. For a single cell, many LSTAs (from thousands of different images) are used to determine a relevant low-dimensional space (by PCA on the gradients, see also (Constantine et al. 2014)) and the dataset is projected on this space. Each point corresponds to a natural image and arrows indicate the direction of the gradient (LSTA), capturing the direction in stimulus space that increases the neural response most. **C.** A visual representation of the gradient field for an ON-OFF cell in the PCA plane. Black arrows show the gradient direction at different points in stimulus space. While MEIs (purple and green trajectories) follow the gradient to local maxima of the response function, different initializations can lead to distinct MEIs, indicating complex, context-dependent tuning. Metamer directions (red arrow) are orthogonal to the gradient and produce invariant responses. This field-based approach reveals how the neuron's selectivity varies across different stimulus contexts and can expose nonlinear computations not captured by a single MEI.

However, MEIs provide information only about the maximum of F and do not characterize the shape of the tuning curve. There are some cases where characterizing the MEI might be insufficient to understand the computations a neuron performs. For example, in Figure 4, we show how a cell that computes spatial contrast will give very different MEIs depending on which starting point in the stimulus space is chosen for the optimization. If an MEI is always the same, no matter the starting point, it might reveal the functional selectivity of the cell. However, it can happen that an MEI changes as a function of the starting point, e.g. for ON-OFF ganglion cells (Höfling et al. 2024). This might be the signature of something more non-linear that needs to be investigated.

To address this issue, a possible option is to study F by sampling its gradient at multiple points in the stimulus space. Rather than focusing solely on peak selectivity, we can aim at describing how neural responses change locally with respect to different stimulus contexts. To this end, we introduced a novel experimental approach—termed the "perturbative approach"—in which we added small perturbations to reference stimuli. We used these data to estimate "local Receptive Fields" (LSTAs) that are data-driven estimates of the gradient of F calculated around each reference stimulus. In this way we determined which changes in the stimulus maximally changed the response of the recorded neuron and which ones left it invariant, when starting from different points in the stimulus space. By estimating local RFs for several different reference stimuli, we were able to estimate empirically the gradient in several points of the stimulus space, and we gained novel insights on how the selectivity of F depends on the stimulation context (Figure 4C). This approach was successful in the mouse retina to identify an RGC type that extracts quadratic contrast from natural scenes (Goldin et al. 2022) and to uncover parallel processing of moving textures across RGC types (unpublished work). A similar approach has been used in the visual cortex to study tuning properties of high-level neurons (Wang and Ponce 2022).

Overall, these different approaches provide ways to understand better retinal processing. They also allow to constrain more qualitatively the models. For example, a linear model can be discarded if the response to the natural stimulus and the linear equivalent are different. A change in local RFs is a signature of non-linear processing: for example, if the local RF (model gradient) changes position when measured for different natural images, a linear model followed by a non-linearity cannot explained this, even if the non-linearity is U-shaped.

These natural stimuli manipulations thus offer a different perspective on how to model retinal processing of natural scenes. Instead of collecting the responses to natural movies and then just learning a model on them, they advocate for a back-and-forth approach where we design stimulus manipulation that should allow to discriminate between classes of models, and then use the results to determine which class of models can be selected or rejected (Figure 5).

All these manipulations have in common to be inspired by mathematical tools that have been used to study theoretical (artificial) systems. Here, these mathematical approaches are directly applied to experimental systems. This "model-free" approach is motivated by the fact that "The real object might constitute the simplest description of itself" (Von Neumann, 1949). Since any model of retinal processing will be an approximation, these direct estimations of the gradients, local maxima, or invariant subspace, are a necessary complement to the mathematical analysis of the models of retinal processing.

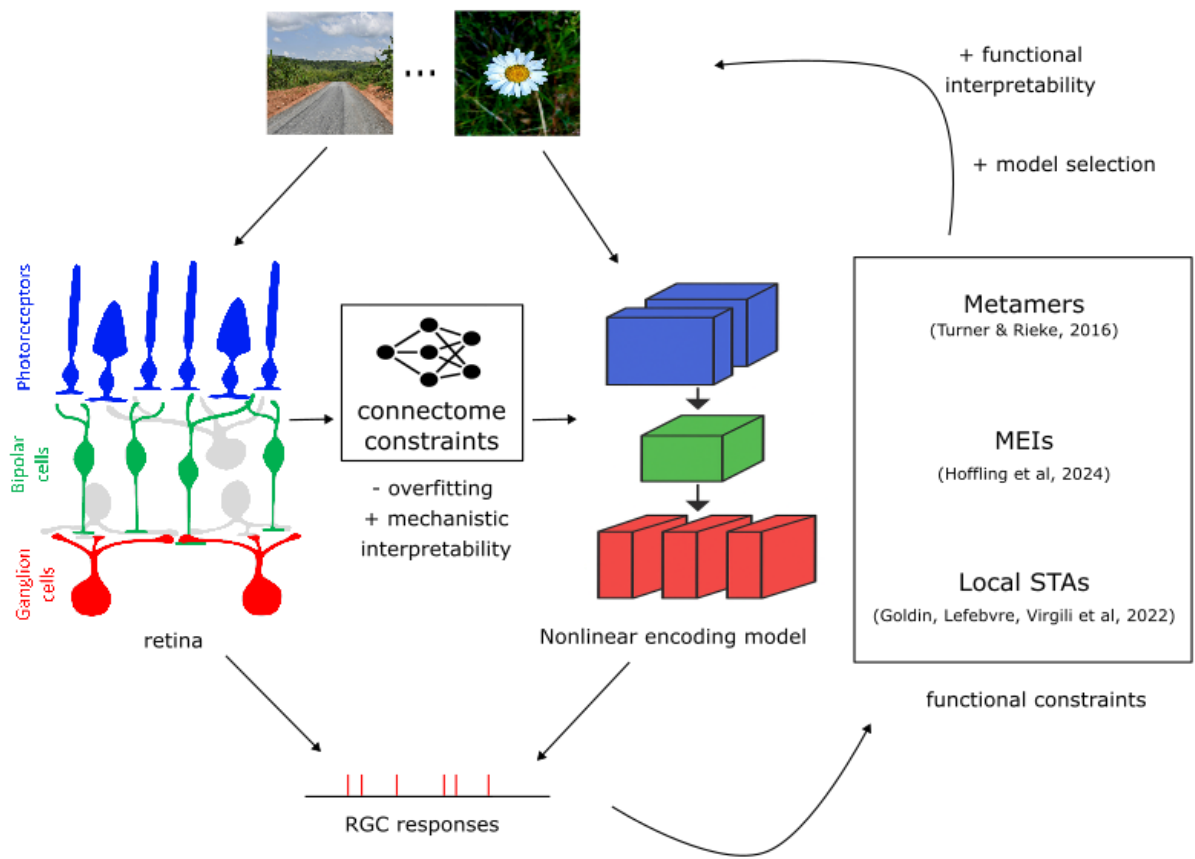

**Figure 5: New structural and functional constraints for encoding approaches**
Due to the complexity of natural stimuli, encoding models of retinal processing under natural stimuli face challenges in interpretability, overfitting, and generalization. Two complementary strategies can help address these issues. **Left path:** Connectomic constraints, such as cell-type specific connectivity derived from structural reconstructions, can be used to constrain model architectures. This reduces the number of free parameters, improves interpretability, and aligns model components with biological circuits. **Right path:** Functional constraints obtained through structured manipulations of natural scenes, such as metamers, MEIs and LSTAs (Figure 4), provide qualitative insights into the computations performed by ganglion cells. These methods can reveal invariance and selectivity guiding model selection.

## 6. Conclusion

To conclude, we think that a different type of reductionism is emerging in the field of retinal computations, and more broadly in sensory neuroscience. The classical reductionist approach was to break down natural stimuli into simpler components (e.g. Fourier components), study the responses to these simpler components, and make hypotheses from there about how natural stimuli will be processed. A new type of reductionism is to break down the broad ensemble of natural stimuli into the different visual inputs corresponding to different natural tasks, and study how the retina processes each of these subsets. This new "divide and conquer" strategy has become possible for several reasons:

1) machine learning approaches allow to have models that predict ganglion cell responses to complex stimuli with a better precision than before.
2) Several ecological visual tasks that animals like mice are able to achieve have been characterized.
3) There are more and more data reconstructing the visual input that falls on the retina during these tasks.

This new reductionism will be instrumental in keeping models of retinal computations interpretable despite their growing complexity. Another way to tame the complexity of the models is to constrain them with biological data. Connectomics are a promising way to reduce the space of possible models. Natural image manipulation is also a way to constrain models and help functional interpretability.

Another emerging trend is to mix different levels of approach. Classically, the way to understand retinal computations can be divided in 3 levels: the normative approaches, which focus on the objective function ; the coding or phenomenological level, which aims at finding models to predict the retinal responses to visual stimuli ; the mechanistic level, that aims at finding the biological mechanisms of specific computations. There are more and more opportunities to mix these levels to make progress, bringing one to the rescue of another.

For example, an objective function may be optimized under the constraint of connectivity derived from connectomic data (Lappalainen et al. 2024). Efficient coding has been used as a prior for the parameters of encoding models (Młynarski et al. 2021). Finally, testing selectivity and invariance with natural stimuli manipulation can help understand the role of specific circuit mechanisms (Chen et al. 2016). Although we focused here mostly on modeling, it is possible that specific circuits will only make sense once they are put in a more natural context, defined by a natural visual task.

**Acknowledgments**: This work was supported by the ERC Consolidator grant DEEPRETINA (101045253) (O.M.), ANR grant Chaire Industrielle MyopiaMaster ANR-22-CHIN-0006 (O.M.), ANR grant ANR-18-CE37-0011–DECORE (O.M.), ANR grant ANR-20-CE37-0018-04–Shooting Star (O.M.), ANR grant ANR-22-CE37-0033 NUTRIACT (O.M.), ANR grant project ANR-22-CE37-0016-01 PerBaCo (O.M.). ANR grant project RetNet4EC (O.M.), AVIESAN-UNADEV (O.M.), Retina France (O.M.), Programme Investissements d'Avenir IHUFOReSIGHT 497 (ANR-18-IAHU-01)

(O.M.). O.M.'s lab is part of the DIM C-BRAINS, funded by the Conseil Régional d'Ile-de-France.